\documentclass[runningheads]{llncs}
\usepackage{cite}
\usepackage{graphicx}
\usepackage{multirow}
\usepackage{amssymb}
\usepackage{amsmath}
\usepackage{bm}
\usepackage[marginal]{footmisc}
\usepackage[misc]{ifsym}
\usepackage[colorlinks,citecolor=blue,urlcolor=blue]{hyperref}
\usepackage{url}

\begin{document}
\title{RPLHR-CT Dataset and Transformer Baseline for Volumetric Super-Resolution from CT Scans}
\titlerunning{RPLHR-CT Dataset and Transformer for Volumetric SR}

\author{
Pengxin Yu \inst{1} \and
Haoyue Zhang \inst{1,2} \and
Han Kang \inst{1} \and
Wen Tang \inst{1} \and
Corey W. Arnold \inst{2} \and
Rongguo Zhang \inst{1}\textsuperscript{(\Letter)}
}
%index{Yu, Pengxin}
%index{Zhang, Haoyue}
%index{Kang, Han}
%index{Tang, Wen}
%index{Arnold, CW}
%index{Zhang, Rongguo}

\authorrunning{Pengxin Yu, Haoyue Zhang et al.}

\institute{
Infervision Medical Technology Co., Ltd. Beijing, China\\ \email{zrongguo@infervision.com}\\
\and
Computational Diagnostics Lab, UCLA, Los Angeles, USA
}
\maketitle
\footnote{Pengxin Yu and Haoyue Zhang contribute equally to this work.}
\begin{abstract}
In clinical practice, anisotropic volumetric medical images with low through-plane resolution are commonly used due to short acquisition time and lower storage cost. 
Nevertheless, the coarse resolution may lead to difficulties in medical diagnosis by either physicians or computer-aided diagnosis algorithms. 
Deep learning-based volumetric super-resolution (SR) methods are feasible ways to improve resolution, with convolutional neural networks (CNN) at their core. 
Despite recent progress, these methods are limited by inherent properties of convolution operators, which ignore content relevance and cannot effectively model long-range dependencies. 
In addition, most of the existing methods use pseudo-paired volumes for training and evaluation, where pseudo low-resolution (LR) volumes are generated by a simple degradation of their high-resolution (HR) counterparts. 
However, the domain gap between pseudo- and real-LR volumes leads to the poor performance of these methods in practice. 
In this paper, we build the first public real-paired dataset RPLHR-CT as a benchmark for volumetric SR, and provide baseline results by re-implementing four state-of-the-art CNN-based methods. 
Considering the inherent shortcoming of CNN, we also propose a transformer volumetric super-resolution network (TVSRN) based on attention mechanisms, dispensing with convolutions entirely. 
This is the first research to use a pure transformer for CT volumetric SR. 
The experimental results show that TVSRN significantly outperforms all baselines on both PSNR and SSIM. 
Moreover, the TVSRN method achieves a better trade-off between the image quality, the number of parameters, and the running time. 
Data and code are available at https://github.com/smilenaxx/RPLHR-CT.

\keywords{Volumetric super-resolution \and CT \and Deep learning \and Transformer}
\end{abstract}

\section{Introduction}
Volumetric medical imaging, such as computed tomography (CT) and magnetic resonance imaging (MRI), is an important tool in diagnostic radiology. Although high-resolution volumetric medical imaging provides more anatomical and functional details that benefit diagnosis \cite{ref3,bgref2,bgref3}, long acquisition time and high storage cost limit the wide application in clinical practice. As a result, it is routine to acquire anisotropic volumes in practice, which have high in-plane resolution and low through-plane resolution. However, the disparity in resolution can lead to several challenges: (1) the inability to display sagittal or coronal views with adequate detail \cite{ref2}; (2) the insufficiency of spatial resolution to observe the details of lesions \cite{ref3} and; (3) the challenge to the robustness of 3D medical image processing algorithms \cite{ref1, ref4}. A feasible solution is to use super-resolution (SR) algorithms \cite{ref5} to upsample anisotropic volumes along the depth dimension, in order to restore high resolution (HR) from low resolution (LR). This approach is referred to as "volumetric SR."

CNN-based algorithms have achieved outstanding performance in SR for natural images \cite{ref6} and these techniques have been introduced for volumetric SR \cite{ref7,ref8,ref9,ref10,ref11,ref12,ref13,ref14,ref15,ref16}. Though significant advances have been made, CNN-based algorithms remain limited by the inherent weaknesses of convolution operators. On the one hand, using the same convolution kernel to restore various regions may neglect the content relevance. Liu et al. \cite{ref8} take this into consideration and propose a multi-stream architecture based on lung segmentation to recover different regions separately, but this is hard to be a one-size-fits-all solution. On the other hand, the non-local content similarity of images has been used as an effective prior in image restoration \cite{ref17}. Unfortunately, the local processing principle of the convolution operator makes algorithms difficult to effectively model long-range dependence. Recently, transformer networks have shown good performance in several visual problems of natural image \cite{ref18,ref19}, including SR \cite{ref20,ref21}. Self-attention mechanism is the key to the success of transformer. Compared to CNN-based algorithms, transformer can model long-range dependence in the input domain and perform dynamic weight aggregation of features to obtain input-specific feature representation enhancement \cite{ref24}. These results prompted us to explore a transformer-based SR method.

Another impediment to the application of volumetric SR methods is data. Most relevant studies use HR volume as ground truth and degrade it to construct paired pseudo-LR volumes with which to train and evaluate methods \cite{ref7,ref10,ref11,ref12,ref14,ref15}. For instance, Peng et al. \cite{ref7} perform sparse sampling on the depth dimension of thin CT to obtain pseudo thick CT. Zhao et al. \cite{ref15} simulate pseudo-LR MRI by applying an ideal low-pass filter to the isotropic T2-weighted MRI followed by an anti-ringing Fermi filter.
However, the performance will be affected when test on the real-LR volume \cite{ref16} because of the domain gap between pseudo- and real-LR volume. To avoid it, some studies collect real-paired LR-HR volumes \cite{ref1,ref8,ref9,ref13,ref16}. For example, Liu et al. \cite{ref8} collect 880 real pairs of chest CTs and construct a progressive upsampling model to reconstruct 1mm CT from 5mm CT. In the field of MRI, a large data set containing 1,611 real pairs of T1-weighted MRIs have been used to develop the proposed SCSRN method \cite{ref13}. However, a benchmark to objectively evaluate various volumetric SR methods is still lacking.

To address this deficiency, the first goal of this work is to curate a medium-sized dataset, named Real-Paired Low- and High-Resolution CT (RPLHR-CT), for volumetric SR. RPLHR-CT contains real-paired thin-CTs (slice thickness 1mm) and thick-CTs (slice thickness 5mm) of 250 patients. To the best of our knowledge, RPLHR-CT is the first benchmark for volumetric SR, which enables method comparison. The other goal of our work is to explore the potential of transformer for volumetric SR. Specifically, we propose a novel Transformer Volumetric Super-Resolution Network (TVSRN). TVSRN is designed as an asymmetric encoder-decoder architecture with transformer layer, without any convolution operations. TVSRN is the first pure transformer used for CT volumetric SR. We re-implement and benchmark state-of-the-art CNN-based volumetric SR algorithms developed for CT and show that our TVSRN outperforms existing algorithms significantly. Additionally, TVSRN achieves a better trade-off between image quality, the number of parameters, and running time.

\section{Dataset and Methodology}
\subsection{RPLHR-CT Dataset}
\noindent \textbf{Dataset Description.} The RPLHR-CT dataset is composed of 250 paired chest CTs from patients. All data have been anonymized to ensure privacy. 
Philips machines were used to perform CT scans and the raw data were then reconstructed to thin CT (1mm) and thick CT (5mm) images.
Thus, recovering thin CT (HR volume) from thick CT (LR volume) for this dataset is a volumetric SR task with an upsampling factor of 5 in the depth dimension.
The CT scans are saved in NIFTI (.nii) format with volume sizes of $L \times 512 \times 512$, where $512 \times 512$ is the size of CT slices, and $L$ is the number of CT slices, ranging from $191$ to $396$ for thin CT and $39$ to $80$ for thick CT. 
The thin CT and the corresponding thick CT have the same in-plane resolution, ranging in $[0.604, 0.795]$, and are aligned according to spatial location.  

\noindent \textbf{Dataset split and Evaluation Metric.}  
We randomly split the RPLHR-CT dataset into 100 train, 50 validation and 100 test CT pairs. 
For evaluation, we quantitatively assess the performance of all methods in terms of peak signal to noise ratio (PSNR) and structural similarity (SSIM) \cite{ref25}. 
Significance is tested by one-sided Wilcoxon signed-rank test.

\begin{figure}
\begin{center}
\includegraphics[width=\textwidth]{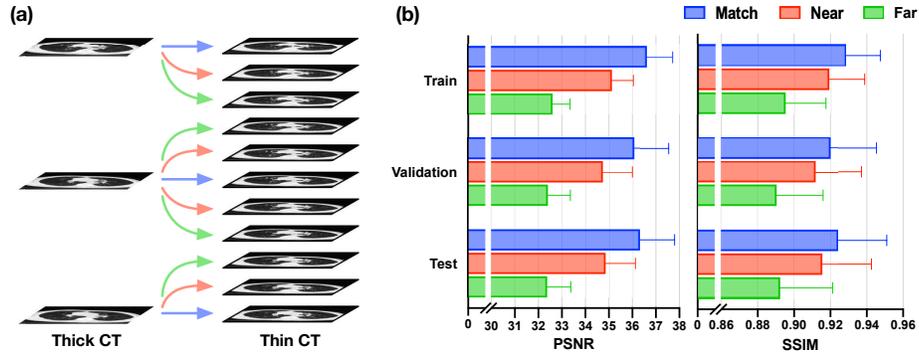}
\end{center}
\caption{(a) Three categories of slice-pairs according to their spatial relationship in thin CT and thick CT. Match: same position, shown in blue; Near: 1mm apart, shown in red; Far: 2mm apart, show in green. (b) The degree of similarity between the three slice-pairs on the three datasets. (Color figure online)} \label{fig1}
\end{figure}

\noindent \textbf{Dataset Analysis.}
To analysis the difference between the thin CT and thick CT, we group slices in thin CT and thick CT into three categories of slice-pairs according to their spatial relationship, as shown on the left side of Fig.~\ref{fig1}. We use PSNR and SSIM to access the changes in the similarity of three slice-pairs in train, validation and test CT pairs. As shown on the right side of Fig.~\ref{fig1}, the results indicate that the similarity of slice-pairs at the same spatial location in thin CT and thick CT, namely \textbf{Match}, is the highest, while the similarity decreases significantly as the spatial distance becomes larger.

\begin{figure}
\begin{center}
\includegraphics[width=\textwidth]{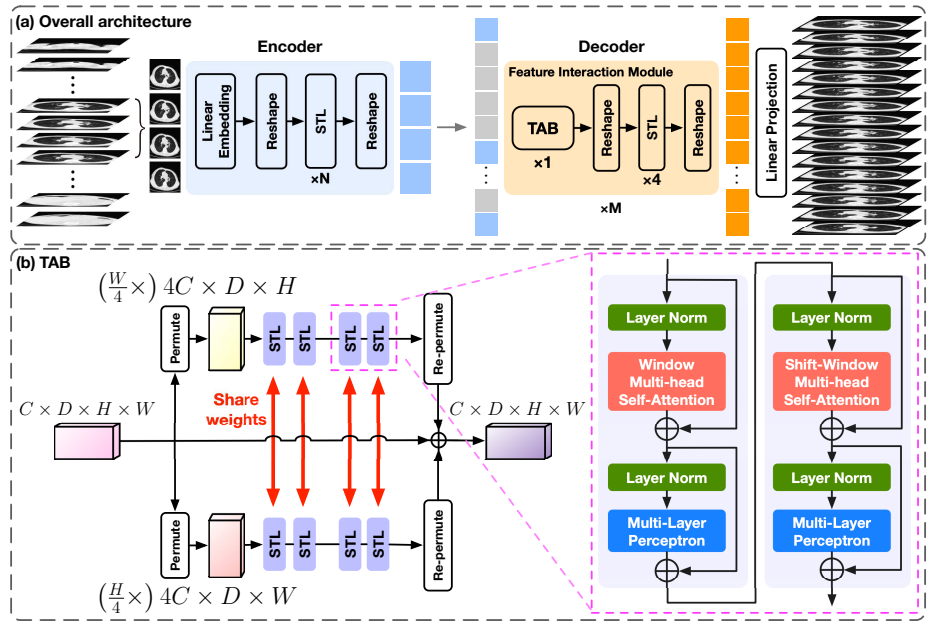}
\end{center}
\caption{(a) Illustration of the proposed Transformer Volumetric Super-Resolution Network architecture. (b) Details of TAB. The purple dashed box represents two consecutive swin transformer layers. The batch dimension is indicated in parentheses.} \label{fig2}
\end{figure}

\subsection{Network Architecture}
Inspired by MAE\cite{ref28}, we treat volumetric SR as a task to recover the masked regions from the visible regions, where the visible regions refer to the slices in the LR volume and the masked regions refer to the slices in the corresponding HR volume. As illustrated in Fig.~\ref{fig2}, we also design our TVSRN with an asymmetric encoder-decoder architecture, but with several targeted modifications. First, in TVSRN, the encoder and the decoder are equally important, and to better model the relationship between the visible regions and the masked regions, the decoder uses a larger amount of parameters than the encoder. Second, instead of the standard transformer layer\cite{ref18}, we use the swin transformer layer (STL)\cite{ref19}, which is less computationally intensive and more suitable for high resolution image, as the basic component of TVSRN. Third, we propose Through-plane Attention Blocks to exploit the spatial positional relationship of volumetric data to achieve better performance.

\noindent \textbf{Encoder} is used to map the LR volume to a latent representation. The consecutive slices from LR volumes are denoted as the input $X_{e}^{in} \in \mathbb{R}^{1 \times D \times H \times W}$ of \textit{encoder}, where $D$, $H$ and $W$ are the depth, height and width, and the channel is $1$.
$X_{e}^{in}$ is firstly fed into the \textit{Linear Embedding}, whose number of feature channel is $C$, to extract shallow features and output $F_{s} \in \mathbb{R}^{C \times D \times H \times W}$. 
Then, $F_{s}$ is reshaped to $F_{0} \in \mathbb{R}^{CD \times H \times W}$. 
We stack $N$ STLs to extract deep features from $F_{0}$ as:
\begin{gather}
F_{i} = H^{STL}_i(F_{i-1}), \;\; i=1,2,...,N
\end{gather}
where $H^{STL}_i(\cdot)$ denotes the $i$-th STL. 
Finally, $F_{N}$ is reshaped to 3D output $X_{e}^{out} \in \mathbb{R}^{C \times D \times H \times W}$.

\noindent \textbf{Decoder} is used to recovery the HR volume from the latent representation. As shown in Fig.~\ref{fig2}(a), mask tokens are introduced after the \textit{encoder}, and the full set of $X_{e}^{out}$ and mask tokens is input to the \textit{decoder} as $X_{d}^{in} \in \mathbb{R}^{C \times D' \times H \times W}$, where $D'$ is the depth of ground truth. 
The mask tokens are learned vector that indicates the missing slices in the LR volumes compared to the HR counterpart. 
\textit{Decoder} stack $M$ Feature Interaction Modules (FIMs), which consists of one Through-plane Attention Block (TAB), four STLs and two reshape operations. 
The reshape operations are used to reshape the input feature map into the size expected by the next block. 
The output of the \textit{decoder} is $X_{d}^{out}$ with the same size as $X_{d}^{in}$. 
Note that the design of asymmetric \textit{decoder} can easily be adapted to other upsampling rates by changing the number of mask tokens.

The details of TAB are illustrated in Fig.~\ref{fig2}(b). TAB is the first block in each FIM. There are two parallel branches in TAB that perform self-attention on the input from coronal and sagittal views, respectively. 
In both views, the depth dimension will become an axis of the STL's window, so the relative position relationship between slices will be incorporated into the calculation. 
The parameter weights of the corresponding STL on the two parallel branches are shared. 
Given the input feature $z_{in}$ of TAB, the output is computed as:
\begin{gather}
z^{sag}_{0} = P^{sag}(z_{in}),\;z^{cor}_{0} = P^{cor}(z_{in}) \notag \\
z^{sag}_{j} = H^{STL}_{j}(z^{sag}_{j-1}),\;z^{cor}_{j} = H^{STL}_{j}(z^{cor}_{j-1}), \;\; j=1,2,3,4 \notag \\
z_{out} = z_{in} + P^{sag}_{re}(z^{sag}_{4}) + P^{cor}_{re}(z^{cor}_{4})
\end{gather}
where $P^{sag}(\cdot)$ and $P^{cor}(\cdot)$ are permutation operations that transform the input to sagittal and coronal view, respectively. 
$P^{sag}_{re}(\cdot)$ and $P^{cor}_{re}(\cdot)$ denote re-permutation operations that reshape the input back to original size.
In addition, TAB contains residual connection, which allow the aggregation of different levels of features.

\noindent \textbf{Reconstruction Target.} 
The $X_{d}^{out}$ is fed into the \textit{Linear Projection} to obtain the pixel-wise prediction $\hat Y \in \mathbb{R}^{D' \times H \times W}$. The $L_1$ pixel loss is formulated as:
\begin{gather}
L_{pixel} = \frac{1}{D' \times H \times W} \sum_{k,i,j}|\hat Y_{k,i,j} - Y_{k,i,j}|
\end{gather}
where $Y$ is the ground truth HR volume.

\noindent \textbf{Architecture Hyper-parameters.} 
For each STL, the patch size is $1 \times 1$ and the window sizes of x-axis, y-axis and z-axis are set to 8, 8 and 4. 
For \textit{Linear Embedding}, the channel number $C$ is 8. 
The number of STLs in encoder and FIMs in decoder is set to $N=4$ and $M=1$, respectively.

\section{Experiments and Results}
\noindent \textbf{Implementation Details.} 
We normalize the intensity of the CT images from $[-1024,2048]$ to $[0,1]$. 
During training, $4 \times 256 \times 256$ cubes from thick CTs are used as input and the corresponding $16 \times 256 \times 256$ cubes from thin CTs are used as ground truth, in where $16 = (4-1) \times 5 + 1$.
During inference, we feed cubes from thick CTs to the model in a sliding window manner, in which the overlap of depth dimension is 1 and the rest is 0.
If the depth of untested cubes is less than $4$, we feed the last $4$ slices into the model. 
For multiple predictions on the same coordinate, we take the average as the final value. 
TVSRN is trained with Adam optimizer. The learning rate is $0.0001$ and the batch size is $1$. 
For the comparison methods, we follow descriptions provided in the original papers to re-implement the models, as none have public code available. 
Settings not detailed in the original paper will remain consistent with our work. 
Data augmentation include random cropping and horizontal flipping. 
The framework is implemented in PyTorch, and trained on NVIDIA A6000 GPUs.  

\begin{figure}
\begin{center}
\includegraphics[width=\textwidth]{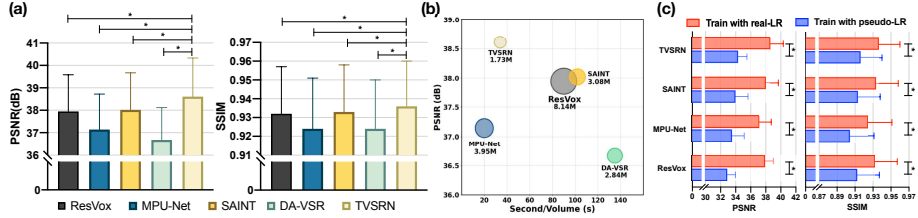}
\end{center}
\caption{(a) Quantitative comparisons of our TVSRN and other state-of-the-art methods. $*$ indicates $p < 0.001$. (b) PSNR vs. processing time of each volume with number of parameters shown in circle size. (c) quantitative results of pseudo images experiment.} \label{fig3}
\end{figure}

\subsection{Results and analysis}   
Fig.~\ref{fig3}(a) summarizes the quantitative comparisons of our method and other state-of-the-art CT volumetric SR methods: ResVox \cite{ref9}, MPU-Net \cite{ref8}, SAINT \cite{ref7} and DA-VSR \cite{ref10}. 
For ResVox, the noise reduction part is removed. 
For MPU-Net, we do not use the multi-stream architecture due to the lack of available lung masks. 
TVSRN achieves PSNR of $38.609\pm1.721$ and SSIM of $0.936\pm0.024$, outperforms others significantly ($p < 0.001$). Moreover, as shown in Fig.~\ref{fig3}(b), compared to other methods, TVSRN achieves a better trade-off in terms of the PSNR (optimal), the number of parameters (optimal), and the running time (suboptimal). We also perform the comparison on an external test set, where TVSRN also achieved the best performance. Detailed numerical results on the internal test set and external test set are presented in the supplementary material. In addition, a sample-by-sample performance scatterplot is given in the supplementary material.     

\begin{figure}
\begin{center}
\includegraphics[width=\textwidth]{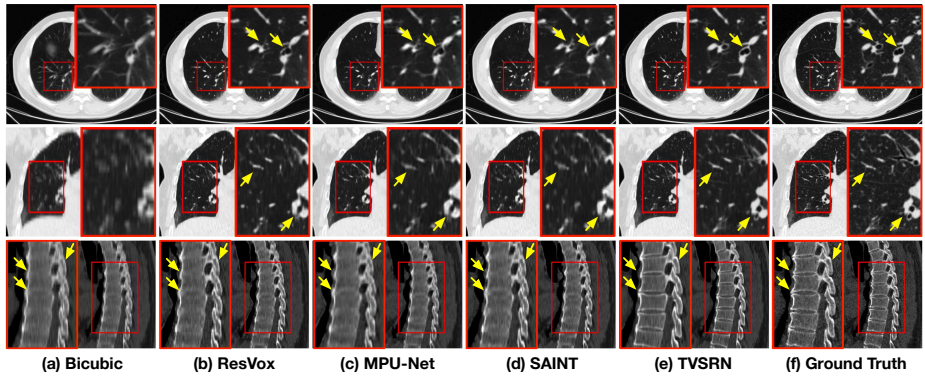}
\end{center}
\caption{Visual comparisons of different methods against TVSRN. The first and second rows show the axial view and coronal view respectively, displayed as lung window. The third row is sagittal view, displayed as bone window. Yellow arrows point to areas of marked difference. } \label{fig4}
\end{figure}

We visualize the axial, coronal and sagittal views of HR CT volume obtained by different methods. It is clear in Fig.~\ref{fig4} that TVSRN has the richest details and the least amount of structural artifacts remaining in different views. 

\subsection{Domain gap analysis}
We conduct a pseudo images experiment to illustrate the effect of the domain gap. Specifically, we degrade the training data to obtain pseudo-LR volumes, and use these data to train several different methods. 
All settings are the same as those in the previous section, except for the training data. 
For testing, real-LR volumes in the internal test set are used as input to calculate the PSNR and SSIM. 
As shown in Fig.~\ref{fig3}(c), the results show that both PSNR and SSIM of various methods are significantly decreased to varying degrees ($p < 0.001$).
Please refer to the supplemental material for more details of degradation.

\subsection{Ablation Study}
The ablation study is used to verify the contribution of each component in TVSRN on performance. The full TVSRN is compared to:
\begin{itemize}
    \item TVSRN$_{ViT}^{Encoder}$. A standard transformer-based method based on \cite{ref18}. We map each patch of size $1 \times 16 \times 16$ to token with length of $512$ and set the number of transformer layers to eight. Instead of asymmetric \textit{decoder}, it uses subpixel conversion \cite{ref26} to perform upsampling.
    \item TVSRN$^{Encoder}$. Only the \textit{encoder} of TVSRN was used. $N$ is increasd to eight and $C$ is increased to 32. The upsampling method is subpixel convert.
    \item TVSRN$^{w/o\; TAB}$. TAB is not used in TVSRN, that is, the relative position relationship among slices is ignored in the network.
\end{itemize}

\begin{table}
\renewcommand\arraystretch{1.2}
\caption{Results of ablation study for TVSRN in terms of PSNR and SSIM. The best results are \textbf{bolded}, and the second best results are \underline{underlined}. * denotes statistically significant ($p < 0.001$) against above method with one-sided Wilcoxon signed-rank test.}\label{tab1}
\centering
\resizebox{\textwidth / 10 * 7}{!}{
\begin{tabular}{l | c | c | c}
\hline
Designs                 & Param  & PSNR($\uparrow$)       & SSIM($\uparrow$) \\ 
\hline
TVSRN$_{ViT}^{Encoder}$ & 17.15M & $35.537 \pm 1.353$     & $0.918 \pm 0.026$ \\ 
TVSRN$^{Encoder}$       & 1.58M & $38.364 \pm 1.675^*$      & $0.934 \pm 0.024^*$ \\
TVSRN$^{w/o\; TAB}$.    & 1.56M & \underline{~$38.497 \pm 1.700^*$~}      & \underline{~$0.935 \pm 0.024^*$~} \\
TVSRN                   & 1.73M & $\bm{38.609 \pm 1.721^*}$ & $\bm{0.936 \pm 0.024^*}$ \\
\hline
\end{tabular}
}
\end{table}

Model performance is summarized in Table.~\ref{tab1}. Notable observations include: 
1) among all designs, TVSRN$_{ViT}^{Encoder}$ has the most parameters but the worst performance, which indicates that it is not feasible to simply apply the transformer to the volumetric SR; 
2) replacing standard transformer layer with STL can greatly reduce the number of parameters and improve the performance by a large margin (up to 2.827dB); 
3) asymmetric decoder can improve performance slightly without changing the number of parameters; 
4) improvements can be seen from TVSRN$^{w/o\; TAB}$ to TVSRN, indicating the effectiveness of modeling the relative position relationship among slices. A sample-by-sample performance scatterplots in supplemental material is used to further illustrate the effectiveness of individual components.

\section{Conclusion}
A persistent problem with volumetric SR is the lack of real-paired data for training and evaluation, which makes it challenging generalize algorithms to real-world datasets for practical applications. 
In this paper, we presented the RPLHR-CT Dataset, which is the first open real-paired dataset for volumetric SR, and provided baseline results by re-implementing four state-of-the-art SR methods. 
We also proposed a convolution-free transformer-based network, which significantly outperformed existing CNN-based methods and has the least number of parameters and the second shortest running time. 
In the future, we will enlarge the RPLHR-CT Dataset and investigate new volumetric SR training strategies, such as semi-supervised learning or using unpaired real data.

~

\noindent \textbf{Acknowledgment.} This work was funded by the National Key Research and Development Project (2021YFC2500703).

\bibliographystyle{splncs04.bst}
\bibliography{ref.bib}
\end{document}